\documentclass[preprint,onecolumn,superscriptaddress,floatfix,showpacs,amsmath,amssymb]{revtex4}

\usepackage{multirow}
\usepackage{graphicx}
\usepackage{bm}
\usepackage{dcolumn}
\usepackage{picture}
\usepackage[flushleft]{threeparttable}
\usepackage{amsmath}
\usepackage{times}
\usepackage{color}
\usepackage{xcolor}
\usepackage{booktabs}
\usepackage{array}
\usepackage{gensymb}
\usepackage[hidelinks]{hyperref}
\usepackage{natbib}
\usepackage{siunitx}
\usepackage{amsmath}
\usepackage{times}
\usepackage{color}
\usepackage{xcolor}
\usepackage{soul}
\usepackage{amsmath}
\usepackage{amssymb}
\usepackage{tabularx}

\newcommand{\cto}{CoTiO$_{3}$}
\newcommand{\nto}{NiTiO$_3$}

\newcommand{\tn}{$T_{\rm N}$}

\newcommand{\mb}{$\mu_{\rm B}$}
\newcommand{\cp}{$c_{\rm p}$}

\newcommand{\cpm}{$c_{\rm p}^{\rm mag}$}
\newcommand{\cpph}{$c_{\rm p}^{\rm ph}$}

\newcommand{\alv}{$\alpha_{\rm V}$}
\newcommand{\alm}{$\alpha_{\rm V}^{\rm {mag}}$}
\newcommand{\alp}{$\alpha^{\rm ph}$}
\newcommand{\alb}{$\alpha_{\rm b}$}
\newcommand{\alc}{$\alpha_{\rm c}$}
\newcommand{\gph}{$\gamma^{\rm ph}$}
\newcommand{\albm}{$\alpha_{\rm b}^{\rm {mag}}$}
\newcommand{\alcm}{$\alpha_{\rm c}^{\rm {mag}}$}

\newcommand{\etal}{\textit{et~al.}}

\begin{document}

\title{Magnetic phase diagram, magnetoelastic coupling and Gr{\"u}neisen scaling in CoTiO$_3$}

\author{M.~Hoffmann \footnotemark } 
\affiliation{Kirchhoff Institute of Physics, Heidelberg University, INF 227, D-69120, Heidelberg, Germany}
  
\author{K.~Dey \footnotemark[\value{footnote}] \footnotetext{both the authors contributed equally}}
\email[email:]{kaustav.dey@kip.uni-heidelberg.de}
\affiliation{Kirchhoff Institute of Physics, Heidelberg University, INF 227, D-69120, Heidelberg, Germany}
	
\author{J.~Werner}
\affiliation{Kirchhoff Institute of Physics, Heidelberg University, INF 227, D-69120, Heidelberg, Germany}

\author{R.~Bag}
\affiliation{Indian Institute of Science Education and Research, Pune, Maharashtra 411008, India}

\author{J.~Kaiser}
\affiliation{Kirchhoff Institute of Physics, Heidelberg University, INF 227, D-69120, Heidelberg, Germany}

\author{H.~Wadepohl}
\affiliation{Institute of Inorganic chemistry, Heidelberg University, 69120, Heidelberg, Germany}

\author{Y.~Skourski}
\affiliation{Dresden High Magnetic Field Laboratory (HLD-EMFL), Helmholtz-Zentrum Dresden Rossendorf, D-01328 Dresden, Germany}
    
\author{M.~Abdel-Hafiez}
\affiliation{Department of Physics and Astronomy, Uppsala University, Box 516, SE-751 20 Uppsala, Sweden}
\affiliation{Physics Department, Faculty of Science, Fayoum University, 63514 El Fayoum, Egypt}
	
\author{S.~Singh}
\affiliation{Indian Institute of Science Education and Research, Pune, Maharashtra 411008, India}
	
\author{R.~Klingeler}
\affiliation{Kirchhoff Institute of Physics, Heidelberg University, INF 227, D-69120, Heidelberg, Germany}
\affiliation{Center for Advance Materials (CAM), Heidelberg University, INF 227, D-69120, Heidelberg, Germany}

\date{\today}

\begin{abstract}
High-quality single crystals of CoTiO$_3$ are grown and used to elucidate in detail structural and magnetostructural effects by means of high-resolution capacitance dilatometry studies in fields up to 15 T which are complemented by specific heat and magnetization measurements. In addition, we refine the single-crystal structure of the ilmenite ($R\bar{3}$) phase. At the antiferromagnetic ordering temperature \tn\, pronounced $\lambda$-shaped anomaly in the thermal expansion coefficients signals shrinking of both the $c$ and $b$ axes, indicating strong magnetoelastic coupling with uniaxial pressure along $c$ yielding six times larger effect on \tn\ than pressure applied in-plane. The hydrostatic pressure dependency derived by means of Grüneisen analysis amounts to $\partial T_{\rm N}/ \partial p\approx 2.7(4)$~K/GPa. The high-field magnetization studies in static and pulsed magnetic fields up to 60~T along with high-field thermal expansion measurements facilitate in constructing the complete anisotropic magnetic phase diagram of \cto. While the results confirm the presence of significant magnetodielectric coupling, our data show that magnetism drives the observed structural, dielectric and magnetic changes both in the short-range ordered regime well-above \tn\ as well as in the long-range magnetically ordered phase.

	\end{abstract}
		
\maketitle

 \section{Introduction}

The recent theoretical proposals on 3$d^7$ Cobalt-based honeycomb magnets as a promising host for Kitaev model physics has sparked enormous interest in these materials \cite{Liu_2018,Sano_2018}. In the context of Kitaev materials, unlike  the conventional 4$d$ and 5$d$ honeycomb ruthenates and iridates, the spin-obit coupling in 3$d$ Co$^{2+}$ ions is comparatively weaker resulting in a combined Kitaev-Heisenberg Hamiltonian with a rich magnetic phase diagram \cite{Sano_2018, Liu_2020}. Following the theoretical proposals, several new experimental materials for example BaCo$_2$(AsO$_4$)$_2$ \cite{Zhong_2020}, Na$_3$Co$_2$SbO$_6$ \cite{Songvilay_2020}, Na$_2$Co$_2$TeO$_6$ \cite{Songvilay_2016,Songvilay_2020, Hong_2021} comprising of Co-based honeycomb structures have been under intense investigation. 

In this report, we investigate \cto\ belonging to the ilmenite titanates family with the general formula $A$TiO$_3$, where $A$ is a 3$d$ transition metal ion. The crystal structure comprises of alternating layers of corner sharing TiO$_6$ and CoO$_6$ octahedra along the $c$ axis. In a particular $ab$ plane, the magnetic Co$^{2+}$ ions are interconnected via the O$^{2-}$ ions and exhibit a buckled honeycomb-like structure~\cite{barth_1934}. The neutron-diffraction studies performed as early as 1964 by Newnham \etal\ \cite{Newnham_1964} and more recently by Elliot \etal\ \cite{elliot2020} on polycrystalline samples, reveal a long-ranged two-sublattice, easy-plane type antiferromagnetic structure below \tn = 38~K~\cite{Balbashov2017,Stickler1967}, with ferromagnetically aligned Co$^{2+}$ spins lying in the $ab$ plane and the layers being coupled antiferromagnetically along the $c$ axis. The easy-plane type magnetic anisotropy is due to a combined effect of crystal-field effects and spin-orbit coupling on high-spin $3d^7$ Co$^{2+}$ ions effectively leading to a pseudospin-1/2 ground state~\cite{Goodenough1967,Lines_1963,Yuan_2019}. 

The recent inelastic neutron scattering (INS) results generated enormous interest in this material. The main features of the magnon dispersion, i.e., the low-energy dispersion of spin waves and the observance of high-energy spin-orbit excitons at 28~meV was captured well with an XXZ-type Hamiltonian~\cite{Yuan_2019,Yuan2020,elliot2020}. Unusual temperature dependence of the spin-orbit excitation was observed below \tn, which was accounted for by some mixing of the ground state doublet and excited state multiplets~\cite{Yuan2020}. Most importantly, the presence of Dirac magnons at symmetry protected points in $k$-space was observed~\cite{Yuan_2019} which makes \cto\ a model system to study non-trivial magnon band topology~\cite{elliot2020}. Application wise, \cto\ has been studied in the past for its adaptability as high-$\kappa$ dielectrics~\cite{Chao_2004,Pan_2001}, resonator antennas~\cite{Ullah_2015} and more recently for its significant magnetodielectric coupling properties~\cite{Harada2016,Dubrovin_2020}. The mechanism yielding the observed significant magnetodielectric coupling in \cto\ is however unclear.  

Keeping in mind the fundamental and technological interest of \cto , we study in detail the high-field magnetization and magneto-structural coupling by means of high-resolution dilatometry. Thereby, we elucidate magnetoelastic coupling and establish the anisotropic magnetic phase diagram. Combination of thermal expansion data with specific heat enables us to analyse the relevant energy scales in terms of a Grüneisen analysis. 

\section{Experimental Methods}

Macroscopic single crystals of \cto\ were grown in a four-mirror optical floating-zone furnace (Crystal System Corporation, Japan) equipped with $4\times150$~W halogen lamps. Phase purity of the powders and pulverized single crystals was studied at room-temperature by means of powder X-ray diffraction (XRD) measurements on a Bruker D8 Advance ECO diffractometer with Cu-K$\alpha$ source. Single crystal X-ray studies were performed at 100~K using an Agilent Technologies Supernova-E CCD 4-circle diffractometer (Mo-K$\alpha$ radiation $\lambda$=0.71073~\AA, micro-focus X-ray tube, multilayer mirror optics). Laue diffraction in the back scattering geometry was performed to study the crystallinity and to orient the single crystals. The composition analysis was performed using scanning electron microscope equipped with energy dispersive x-ray (EDX) analysis (Zeiss ultra plus). 

Static magnetic susceptibility $\chi=M/B$ was studied in magnetic fields up to 15~T applied along the principal crystallographic axes by means of a home-built vibrating sample magnetometer~\cite{klingeler_2006} (VSM) and in fields up to 5~T in a Quantum Design MPMS-XL5 SQUID magnetometer. The angular dependence of magnetisation was measured in a Quantum Design MPMS3 SQUID magnetometer using a horizontal rotator.
Pulsed-magnetic-field magnetization was studied up to 60~T at the Helmholtz Zentrum Dresden Rossendorf by an induction method using a coaxial pick-up coil system~\cite{Skourski_2011}. The pulse raising time was 7~ms. The pulsed-field magnetization data were calibrated using static magnetic field measurements. Specific heat measurements have been done in a Quantum Design PPMS using a relaxation method. The relative length changes $dL_i/L_i$ were studied on a single crystal of approximate dimensions $1.6 \times 2.0 \times 1.3~$mm$^{3}$ by means of a standard three-terminal high-resolution capacitance dilatometer ~\cite{Kuechler2012,Johannes_2017}. The measurements were performed in magnetic fields up to 15~T and the uniaxial thermal expansion coefficients $\alpha_i = 1/L_i\cdot dL_i(T)/dT$ were derived from the data.

\section{Crystal growth and characterization}
 
 \cto\ powders were synthesized by mixing stoichiometric amounts of Co$_3$O$_4$(99.9 \% Alfa Aesar) and TiO$_2$(99.9 \% Sigma-Aldrich) and sintering in air, at temperatures ranging from 900 to 1150$^\circ$~C with several intermediate grinding steps, until a single phase material was achieved. For crystal growth, about 7-9~cm of homogeneous and dense rods were obtained by first hydrostatically pressing the phase pure \cto\ powders in rubber tubes and then sintering them at 1350$^\circ$~C for 24~h. Owing to its congruently melting nature~\cite{BREZNY1969649}, higher growth rates of 6-8~mm/h were employed for crystal growth. This is in contrast to other incongruently melting ilmenites, for example \nto\ where a slower rate of 3~mm/h must be employed to successfully grow macroscopic single crystals~\cite{Dey2020}. Two growth experiments in differing gas flows comprising of air at ambient pressure and O$_2$ at 1~bar were performed. Both the growths were relatively stable and resulted in macroscopic single crystals. The optimized growth parameters are listed in table~\ref{table1}.    

 
 \begin{table}[htb]
\centering
\caption{Optimized growth parameters, lattice parameters, and phase analysis from the Rietveld refinement of the room temperature powder XRD data of crushed \cto\ single crystals of two growth experiments (I and II). Feed and seed rods were counter-rotated at the same rotation speed.}\vspace{1mm}
\begin{tabularx}{0.8\columnwidth} {>{\raggedright\arraybackslash}X  >{\centering\arraybackslash}X  >{\centering\arraybackslash}X  }
\hline\hline
            &    I       &       II       \\
\hline
   atmosphere             &   air         &       O$_2$     \\
   pressure               &  ambient      &       1~bar     \\
growth-rate (mm/h)        &   6           &        6        \\
rot. speed (mm/h)         &   20          &        20        \\
latt. parameter $a$ (\AA) &   5.066(7)    &       5.065(3)    \\
latt. parameter $c$ (\AA) &  13.918(7)    &     13.916(2)     \\
composition (Co : Ti)     &   1 : 1.004   &      1 : 1.08     \\
secondary phase           &   TiO${_2}$ + Co$_2$TiO$_4$   &     TiO${_2}$ +Co$_2$TiO$_4$         \\	
\hline\hline
\end{tabularx}
\label{table1}
\end{table}

 
\begin{figure}[tbh]
\centering
\includegraphics[width=0.8\columnwidth]{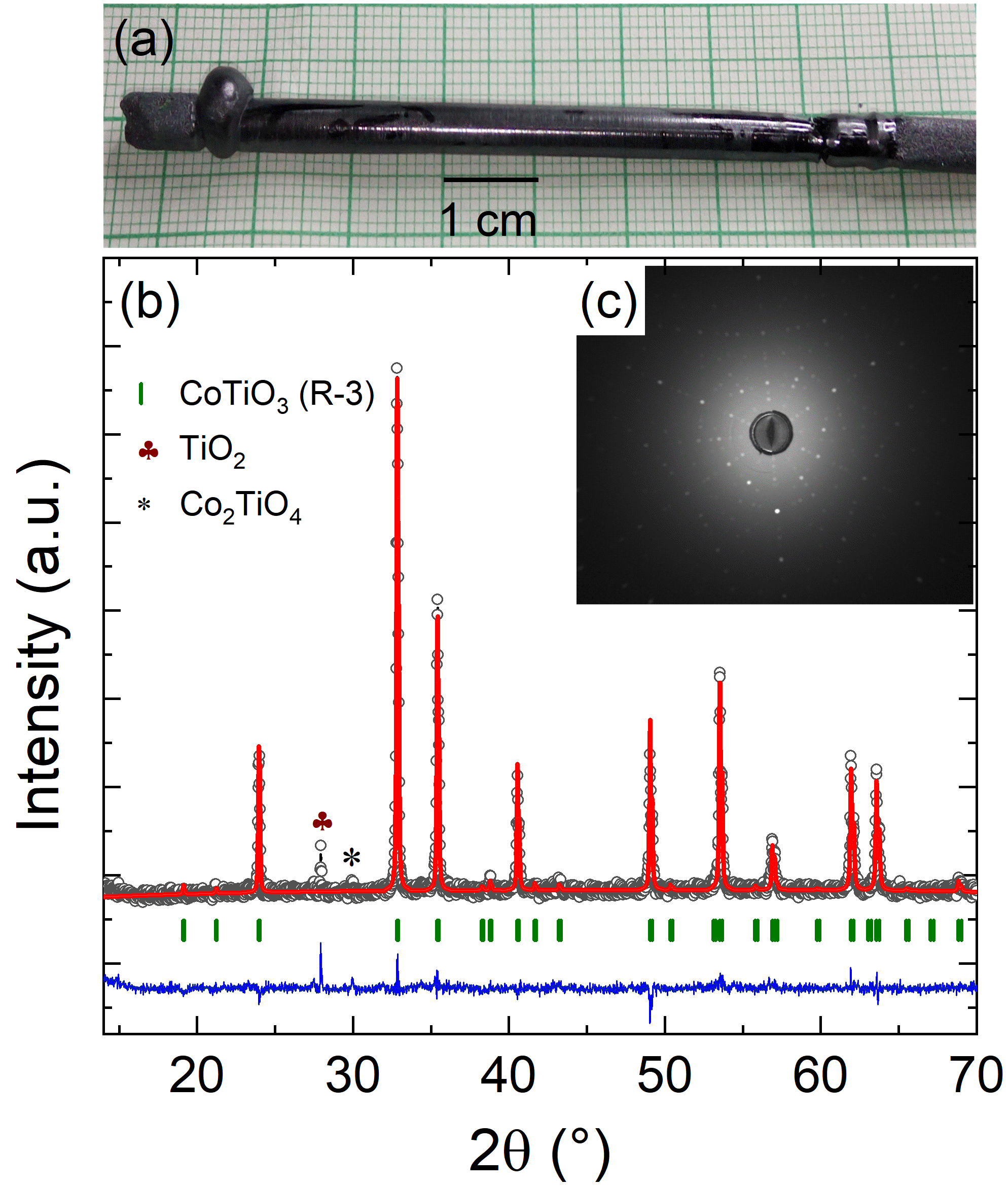}
\caption {(a) Picture of an as-grown \cto\ boule obtained in O$_2$ atmosphere at 1~bar pressure. (b) Rietveld refinement fit of the room temperature XRD data of a powdered \cto\ single crystal. The observed diffraction pattern is shown in black, calculated pattern in red and the difference between observed and calculated pattern is shown in blue. The vertical ticks in green denote the allowed Bragg positions of the ilmenite phase and (c) is a representative Laue pattern of \cto\ single crystal.  }
\label{xrd}
\end{figure}


Fig.~\ref{xrd}(a) shows a representative \cto\ boule grown in O$_2$ atmosphere at 1~bar pressure. The phase analysis by means of Rietveld refinements of powder XRD data of the pulverized single crystalline pieces (see Fig.~\ref{xrd}(b) and Table~\ref{table1}) implies the main ilmenite ($R\bar{3}$) phase as well as the secondary TiO${_2}$ (Rutile) and the spinel (Co$_2$TiO$_4$) phase in both air and oxygen atmosphere grown crystals. Such secondary phases were observed as inclusions in previously reported \cto\ crystals by Balbashov \etal~\cite{Balbashov2017} and ascribed to high-temperature gradients at the growth interface. A qualitative comparison of the relative intensities of impurity peaks from room temperature XRD data reveals that the spinel Co$_2$TiO$_4$ phase is present in larger proportions in air-grown than in oxygen-grown single crystals (where it is hardly visible; see Fig.~\ref{xrd}(b)) in agreement with Balbashov~\etal. However, the EDX analysis suggests a better stoichiometric proportion of Co:Ti for air-grown as compared to oxygen-grown crystals. We ascribe this to a slightly increased evaporation of cobalt oxides in oxygen flow as compared to air flow resulting in stoichiometric mismatch which furthermore leads to a higher proportion of precipitated TiO$_2$ phase in oxygen-grown crystals (see Fig.~\ref{xrd}(b)). However, as will be shown below, the sharp anomalies in magnetization and thermodynamic measurements imply negligible effects of the minor stoichiometric mismatch. Importantly, the back-scattered Laue diffraction spots for the oxygen-grown samples are sharp indicating a high-quality (Fig.\ref{xrd}(c)) as compared to air-grown crystals where the splitting of spots was observed at various points along the grown boule. Due to relatively low proportions of magnetic spinel impurity and better quality, we employed cut and oriented \cto\ single crystals grown in O$_2$ flow at 1~bar for further studies. 

To the best of our knowledge, the earlier studies of ilmenite-type \cto\ crystal structure have been limited to powder diffraction experiments only~\cite{barth_1934,Newnham_1964,Balbashov2017}. We have re-investigated the crystal structure of our oxygen-grown single crystals by means of high resolution single crystal XRD at 100~K with Mo K$\alpha$ radiation ($\lambda$ = 0.71073~\AA). Experimental and refinement details are given in the supplementary material. Similar to the recent single-crystal X-ray diffraction study on \nto \cite{dey_2021}, three somewhat different models were employed for the atomic structure factors $f_{at}$ within the independent spherical atoms approximation: conventional $f_{at}$ calculated with neutral atoms for Co, Ti and O (model A) and two “ionic” models\cite{Angel_2016} ($f_{at}$ for  Co$^{2+}$, Ti$^{4+}$ and O$^{2-}$ (models B and C); for details see  SI). The different models refined to essentially the same structure, with a marginally better fit of model B, but only insignificant differences in the key parameters like atom coordinates, $R$ values, $U_{eq}$ for all atoms and residual electron density. Inter-atomic distances agreed within one standard deviation. The structural refinements confirm the assignment of $R\bar{3}$ space group and improve the accuracy of the crystallographic parameters previously obtained. The obtained lattice parameters and relevant crystallographic information are listed in table~\ref{tab_SXRD} \cite{ICSD}.

\begin{table}[htb]
	\centering
	\caption{Fractional atomic coordinates, Wyckoff positions, site occupation and equivalent isotropic displacement parameters (\AA$^2$) of \cto\ as obtained from refinement of single-crystal XRD measurements at 100~K using model B (see text for more details). [Space group: $R\bar{3}$ (148), $a$ = $b$ = 5.0601~\AA, $c$ = 13.8918~\AA, $\alpha$ = $\beta$ = 90$^{\circ}$, $\gamma$ = 120$^{\circ}$; $sof$ denotes the fraction of atom type present at the site after application of crystallographic symmetry;}\vspace{1mm}
	\begin{tabularx}{1\columnwidth} {>{\raggedright\arraybackslash}X  >{\centering\arraybackslash}X  >{\centering\arraybackslash}X>{\raggedright\arraybackslash}X>{\raggedright\arraybackslash}X>{\raggedright\arraybackslash}X>{\raggedright\arraybackslash}X  }
		\hline\hline
		Atom & Site & x & y & z & $sof$ & U$_{eq}$$^a$ \\ 
		\hline
     	Co &  6c &   0  &  0 &  0.14511(2) &  1 &   0.00297(2) \\
	    Ti &  6c   &      0         &  0	      &  0.35448(2)   &  1   &  0.00262(2) \\
     	O  &  18f  &  0.02051(6)   &  0.31608(6) &  0.25410(2)   &  1   &  0.00391(3) \\
		\hline\hline
	\end{tabularx}
	\label{table1}
	\vspace{1mm}
	\begin{tablenotes}
		\item Note: $^a$ U$_{eq}$ is defined as one third of the trace of the orthogonalized $U_{ij}$ tensor. The anisotropic displacement factor exponent takes the form: $-2 \pi^2[h^2a^{*2}U_{11} + ... + 2hka^*b^*U_{12}]$.]
	\end{tablenotes}
\end{table}


\section{Magnetoelastic coupling}


\begin{figure}[htb]
 \centering
 \includegraphics[width=0.8\columnwidth]{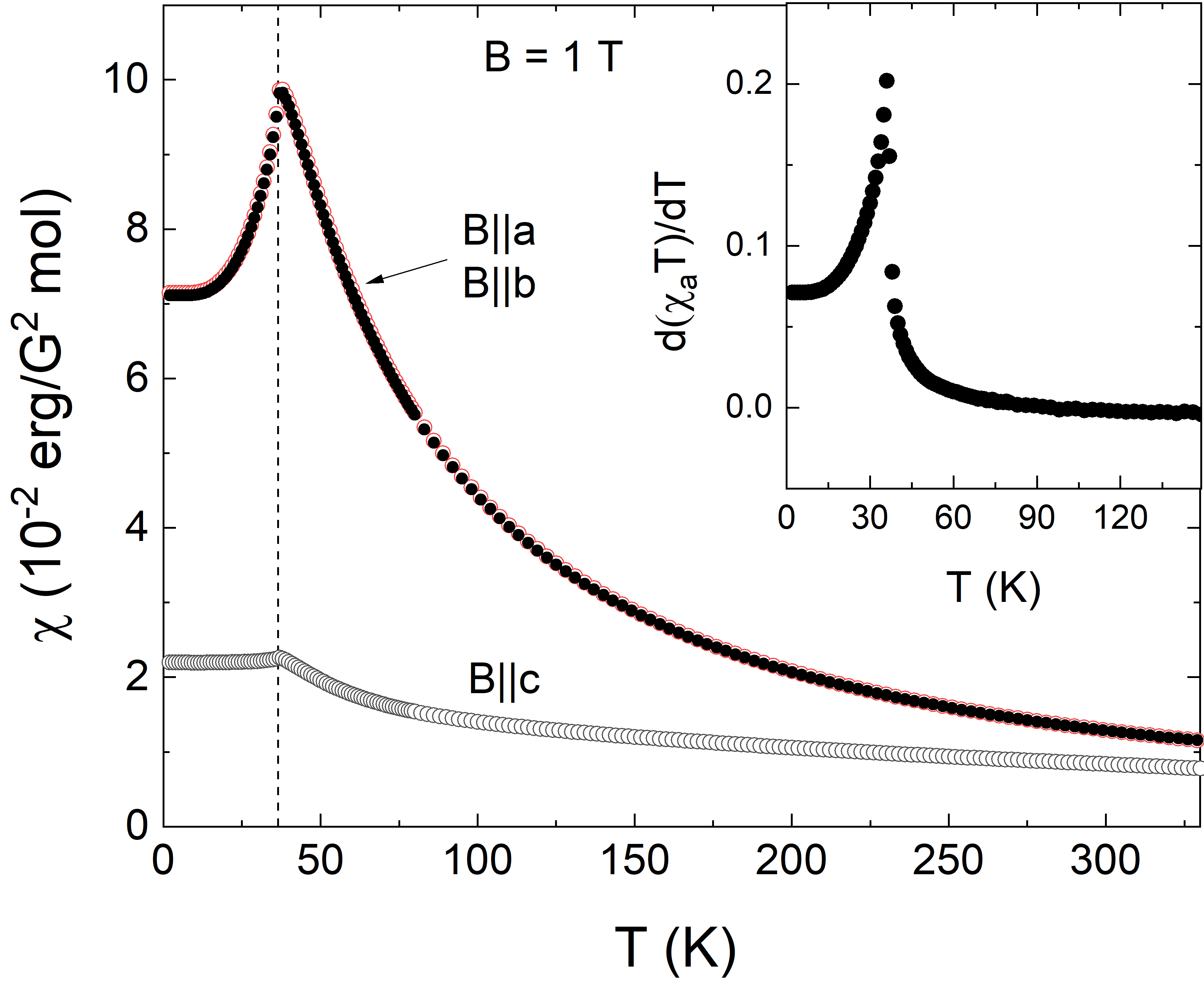}
 \caption {Temperature dependence of static magnetic susceptibility $\chi = M/B$ at $B$ = 1~T applied along the principal crystallographic directions. The inset presents Fisher's specific heat~\cite{Fisher_1962} $\partial(\chi T)/\partial T$ at $B||a$=1~T.}
 \label{fig0_chis}
 \end{figure}
 
\begin{figure}[h]
\centering
\includegraphics[width=0.7\columnwidth]{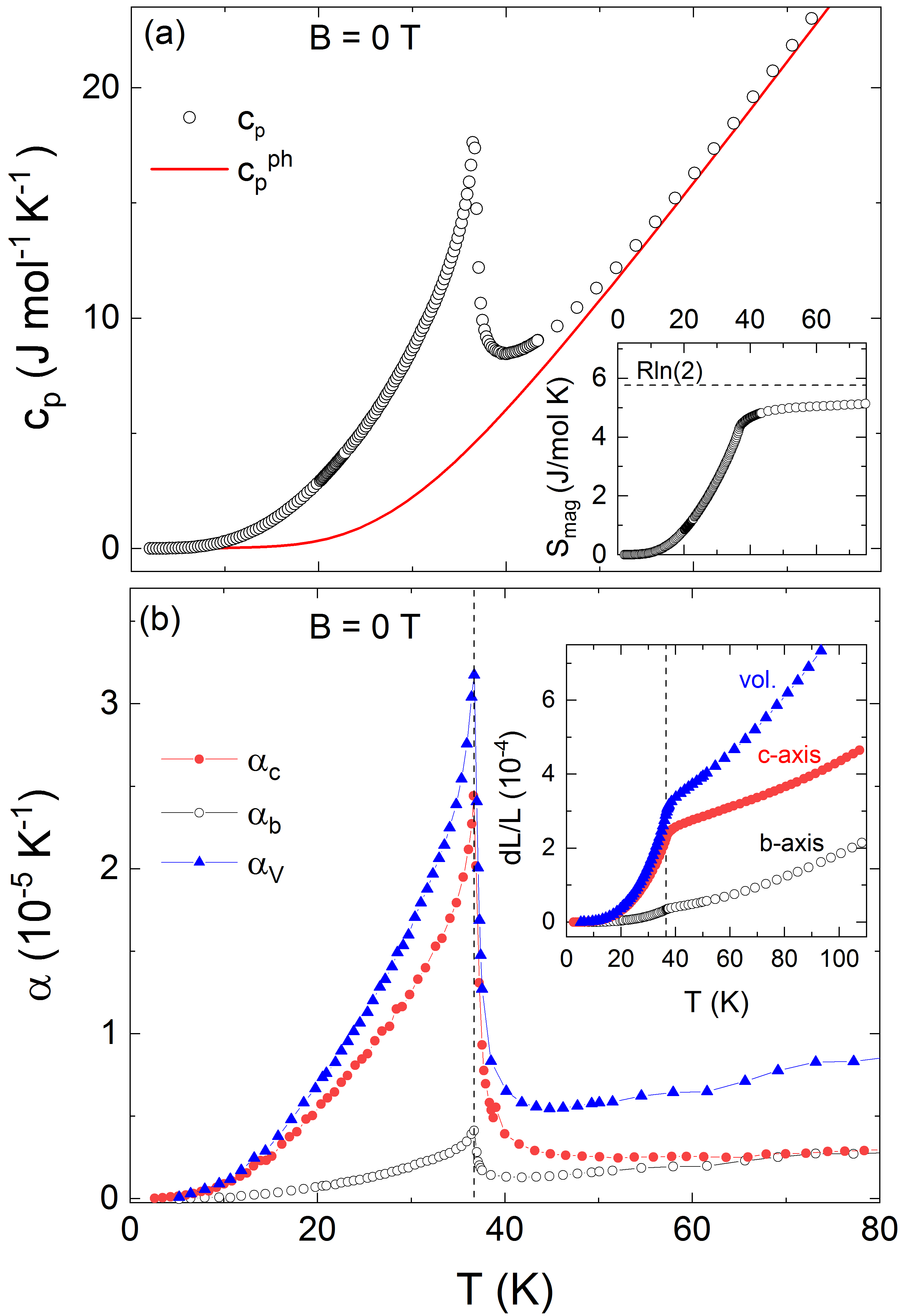}
\caption {(a) Temperature dependence of specific heat $c_{\rm p}$ in zero magnetic field. The solid red line in (a) indicates the phonon specific heat \cpph\ obtained by fitting the \cp\ data with a combined Debye and Einstein model well above the magnetic ordering transition (see the text). The inset to (a) shows the magnetic entropy changes obtained by integrating (\cp\ - \cpph)/$T$. (b) Thermal expansion coefficients $\alpha_i$ versus temperature along the crystallographic $b$ and $c$ axes and the volume thermal expansion coefficient $\alpha_{\rm V}$. The dashed vertical line marks \tn. The inset to (b) shows the associated relative length changes $dL_i/L_i$ versus temperature.}
\label{fig:alpha_cp}
\end{figure}

 
The onset of long-range antiferromagnetic order in \cto\ at \tn\ = 37~K is associated with pronounced anomalies in the magnetic susceptibility (Fig.~\ref{fig0_chis}) as well as in the specific heat and the thermal expansion (Fig.~\ref{fig:alpha_cp}). The magnetic susceptibility is nearly isotropic in the $ab$ plane and exhibits a significant anisotropy with respect to the $c$ axis up to the highest measured temperatures of 350~K. For $T\leq T_{\rm N}$, the susceptibility decreases for magnetic fields $B$ applied in the $ab$ plane and attains a constant value for $B || c$ axis suggesting an easy-plane-type antiferromagnetic structure. This is in accordance with previous studies on polycrystalline~\cite{Newnham_1964,Stickler1967} and single crystalline~\cto\ samples~\cite{Balbashov2017,Watanabe_1980}. Above \tn, the persisting anisotropy up to 350~K is  attributed to single-ion effects due to the octahedral crystal field and spin-orbit coupling on magnetic Co$^{2+}$ spins which results in an effective spin $J_{\rm eff}$ = 1/2 ground state~\cite{Lines_1963,Yuan_2019}. The high-temperature susceptibility behaviour cannot be accounted for by the Curie-Weiss model with an anisotropic $g$-factor. This is explained by considerable mixing of the ground state $J_{\rm eff}$ = 1/2 with excited state multiplets, i.e., $J_{\rm eff}$ = 3/2, 5/2... as experimentally observed in recent INS studies on \cto~\cite{Yuan_2019,Yuan2020,elliot2020} and theoretically described by Goodenough~\cite{Goodenough1967}. In the literature, the high-temperature susceptibility of cobaltates comprising of Co$^{2+}$ ions in octahedral crystalline fields is reported to show complicated temperature dependencies for example in CoCl$_2$~\cite{Lines_1963} and Na$_3$Co$_2$SbO$_6$~\cite{Liu_2020} among several others.


The sharp $\lambda$-shaped anomalies in the specific heat [Fig.~\ref{fig:alpha_cp}(a)] and in Fisher's specific heat~\cite{Fisher_1962} $\partial (\chi_a T)/\partial T$ (Fig.~\ref{fig0_chis} (inset)) confirm the onset of long-range order at \tn\ and also the continuous nature of the phase transition. In order to assess the magnetic entropy changes, the phononic contribution to the specific heat (\cpph) has been estimated by fitting the previously published \cp\ data from Klemme \etal~\cite{Klemme_2011} at temperatures well above \tn\ by an extended Debye model which includes both Debye and Einstein terms~\cite{ashcroft_mermin}. The model fits very well the data at temperatures above 70~K and yields characteristic Debye and Einstein temperatures of $\Theta_D =626(20)$~K and $\Theta_E = 193(10)$~K, respectively. Integrating the magnetic specific heat (\cp - \cpph)/$T$ yields the magnetic entropy changes $S_{\rm mag} = 5.2(2)$~J/(mol K) which is close to the theoretically predicted value of R$ln(2) = 5.7$~J/(mol K) for $J_{\rm eff}$ = 1/2 Co$^{2+}$ spins. The results imply that approximately 15~$\%$ of the total magnetic entropy is consumed between \tn\ and 70~K, suggesting the presence of considerable short-range correlations precursing the evolution of long-range magnetic order.

\begin{figure}[h]
\centering
\includegraphics[width=0.8 \columnwidth]{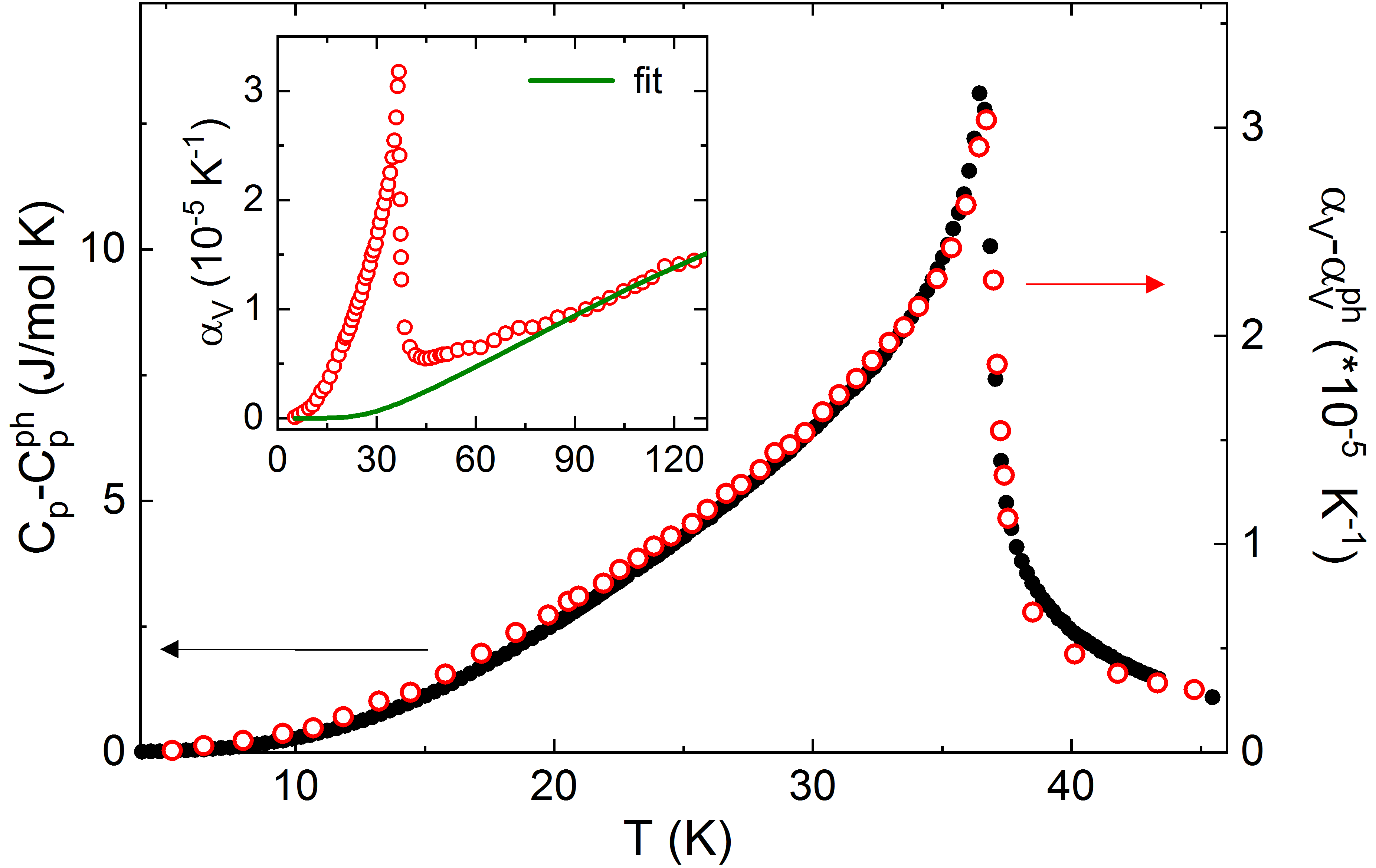}
\caption {Grüneisen scaling of the magnetic contributions to the heat capacity \cpm\ and volume thermal expansion coefficient \alm. Inset: The volume thermal expansion coefficient \alv\ along with a combined Debye-Einstein fit to the high-temperature data (see the text).}
\label{fig:grueneisen_plot}
\end{figure}

The zero-field thermal expansion measurements reveal strong anomalies at \tn\ in the uniaxial thermal expansion coefficients $\alpha_i (i = b,c)$ and in the relative length changes $dL_i/L_i$ [Fig.~\ref{fig:alpha_cp}(b)]. The anomalies demonstrate large spontaneous magnetostriction at \tn\ and hence the presence of significant magnetoelastic coupling in \cto. The measured relative length changes shown in the inset to Fig.~\ref{fig:alpha_cp}(b) signal shrinking of both $c$ and $b$ axes upon the evolution of magnetic order at \tn\ with the size of the anomaly in \alc\ about six times larger than in \alb. As these data imply positive uniaxial pressure dependencies both for pressure applied in-plane and along $c$, the anomaly in the volume thermal expansion coefficient \alv = \alc\ + 2\alb\ correspondingly signals a significant positive hydrostatic pressure dependency. Furthermore, \alb\ and \alc\ in Fig.~\ref{fig:alpha_cp}(b) (also see inset to Fig. \ref{fig:grueneisen_plot}) evidence structural effects above \tn\ precursing the onset of long-range order up to around 70~K. This coincides with the temperature regime where magnetic entropy changes mark the presence of short-range magnetic correlations. 

Comparison of the non-phononic contributions to the thermal expansion coefficient and the  specific heat enables further conclusions on the nature of the associated entropy changes and on magnetoelastic coupling. Exploiting Grüneisen scaling for the lattice, we have approximated the phononic contribution to the volume thermal expansion coefficient \alv\ by using the same $\Theta_{\rm{D}}$ and $\Theta_{\rm{E}}$ from the background specific heat \cpph\ (Fig.~\ref{fig:alpha_cp}(a)) and scaling the Debye and Einstein contributions accordingly~\cite{Dey2020}. As seen in the inset to Fig.~\ref{fig:grueneisen_plot}, the background to \alv\ is well approximated above 90~K. This procedure yields the lattice Grüneisen parameters \gph = \alp/\cpph ~\cite{klingeler_2006} which amount to $\gamma_{\rm{D}} = 1.02 \times 10^{-7}$~mol/J and $\gamma_{\rm{E}} = 1.0(3)\times 10^{-7}$~mol/J, respectively. 

The obtained non-phononic contributions to the thermal expansion coefficient \alm\ are shown with the non-phononic specific heat \cpm\ in Fig.~\ref{fig:grueneisen_plot}(a). Both quantities vary proportionately in the entire $T$-range, i.e., the magnetic Grüneisen parameter is $T$-independent. This observation implies the presence of a single dominant energy scale $J$~\cite{Gegenwart_2016,klingeler_2006}. Since the entropy changes are of magnetic nature, we conclude that a single dominant magnetic degree of freedom drives the observed non-phononic length and entropy changes. The corresponding magnetic volume Grüneisen parameter amounts to $\Gamma^m_{\rm V}$ = \alm/\cpm = 24(2) $\times 10^{-7}$~ mol/J. The hydrostatic pressure ($p_h$) dependency of \tn\ is obtained from the Ehrenfest relation $\partial$\tn/$\partial p_{\rm h}$ = \tn$V_{\rm m}\Gamma_V^m = 2.7(4)$~K/GPa. Here, we used the molar volume $V_{\rm m} = 3.09 \times 10^{-5}$~m$^3$/mol. Furthermore, Grüneisen scaling for each individual axis is confirmed by good proportionality between the uniaxial thermal expansion coefficients \albm\ and \alcm\ and \cpm\ (see the SI Fig. 2) from which we read-off $\Gamma_c^m$ = 1.8(4) $\times 10^{-6}$~ mol/J and $\Gamma_b^m$ = 3(1) $\times 10^{-7}$~ mol/J, respectively. This yields the uniaxial pressure dependencies of $\partial$\tn/$\partial p_c$ = $2.1(5)$~K/GPa and $\partial$\tn/$\partial p_b$ = $0.3(1)$~K/GPa, respectively. 

\section{High-field magnetization and the phase diagram}

\begin{figure}[h]
 \centering
 \includegraphics[width=0.75\columnwidth]{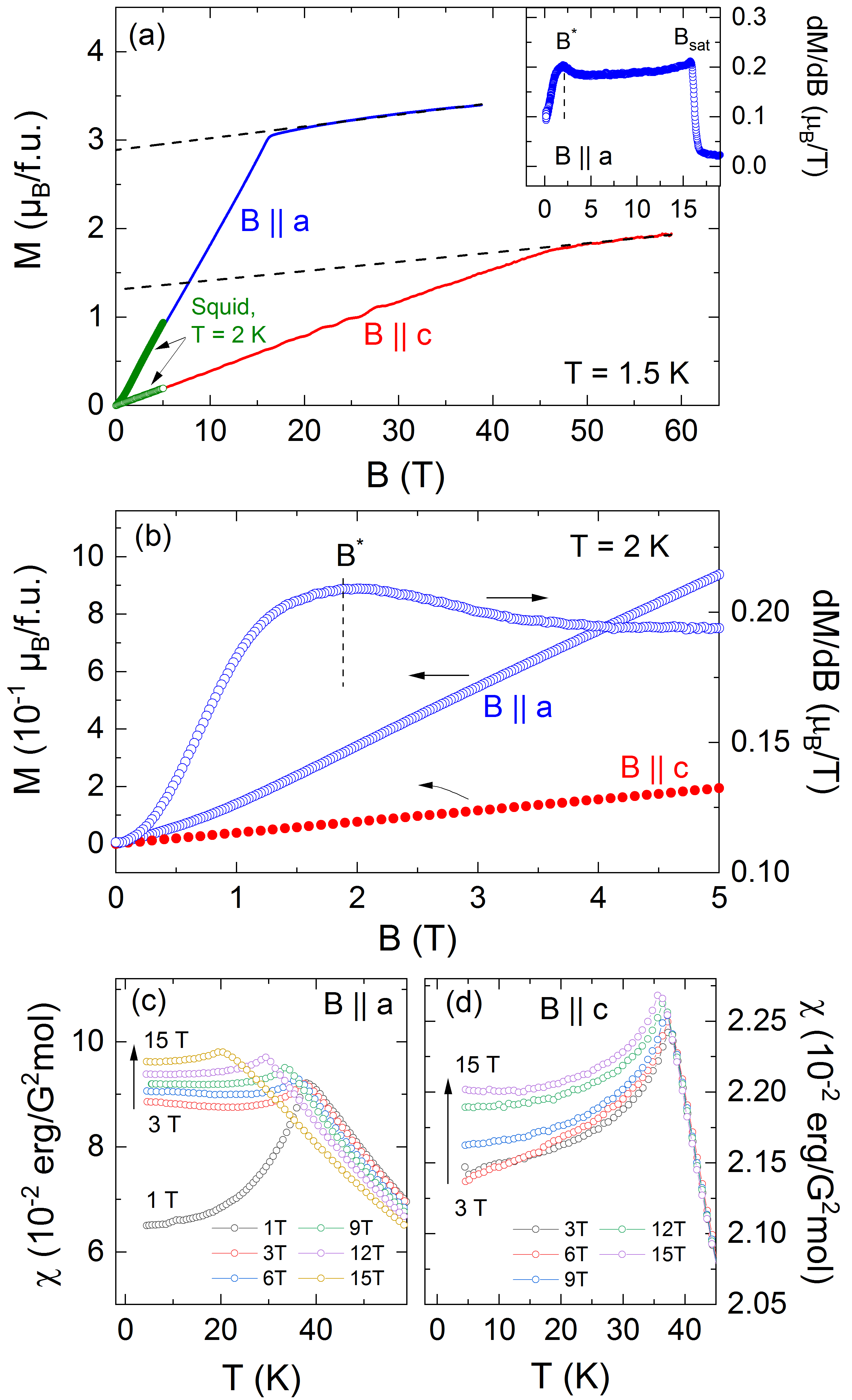}
 \caption {(a) Pulsed-field magnetization $M$ at $T$ = 1.5~K. Dashed lines in (a) denote Van Vleck paramagnetism. The inset to (a) shows the field derivative $\partial M/\partial B$ for $B || a$. (b) Quasi-static field magnetization $M$ and magnetic susceptibility $\partial M/\partial B$ versus magnetic field along the $a$ and $c$ axes, at $T$ = 2~K. (c) and (d) Static magnetic susceptibility $\chi = M/B$ versus temperature for magnetic fields up to 15~T applied along the $a$- and $c$ axes, respectively.}
 \label{fig:PM_B_Chis}
 \end{figure}

The saturation fields and moments at $T = 1.5$~K are determined from pulsed-field magnetization studies up to 60~T as shown in Fig.~\ref{fig:PM_B_Chis}(a). The absolute values of magnetization are calibrated with the SQUID data shown in Fig.~\ref{fig:PM_B_Chis}(a,b). The magnetization is anisotropic and for both $B||a$ and $B||c$ shows largely linear behaviour in a wide range of applied magnetic fields. At any particular field, the magnetization is higher for $B||a$ as compared to $B||c$ which is expected for a highly anisotropic easy-plane antiferromagnet. The saturation fields are determined from the peaks in $\partial M/ \partial B$ (Fig.~\ref{fig:PM_B_Chis}(a) inset), and we obtain $B_{\rm sat,ab} = 16.3(5)$~T and $B_{\rm sat,c} = 46(1)$~T for fields in the $ab$ plane and along the $c$ axis, respectively. Above $B_{\rm sat}$, the magnetization increases linearly which is ascribed to Van Vleck paramagnetism of the Co$^{2+}$ ions in an octahedral environment. From the magnetization slope determined from linear-fits (dashed lines in Fig.~\ref{fig:PM_B_Chis}(a)) above the saturation fields we obtain the Van Vleck susceptibility as $\chi^{\rm VV}_{\rm ab} = 0.013$ erg/(G$^2$mol) and $\chi^{\rm VV}_{\rm c} = 0.011$ erg/(G$^2$mol), respectively. These values are similar to those in other Co-based systems~\cite{Yin_2019,suzuki_2013}. The saturation magnetisation and corresponding $g$-factors obtained after appropriate Van Vleck correction, amounts to $M_{\rm sat,ab} = 2.89$~\mb/f.u. and $g_{\rm ab} = 5.7(2)$ for the $ab$ plane and $M_{\rm sat,c} = 1.31$~\mb/f.u. and $g_{\rm c} = 2.62(4)$ for the $c$ axis, respectively. 

A closer look at the low-field behaviour as shown in Fig.~\ref{fig:PM_B_Chis}(b) at $T = 2$~K, confirms the linearity of $M$ for $B||c$ extending to zero magnetic field whereas a non-linear behaviour (sickle-shaped) up to 4~T is observed for magnetic fields applied in the $ab$ plane. Specifically, the derivative of magnetization with respect to magnetic field shows a broad peak centered at $B^* = 2$~T suggestive of a spin-reorientation process and previously described as a continuous rotation of Co$^{2+}$ moments in the basal plane aided by magnetic field~\cite{Balbashov2017}. Increasing temperature has negligible effects on $B^*$ (see SI Fig. 3) resulting in a horizontal phase boundary (Fig.~\ref{fig:PD}). Static magnetic susceptibility measured in magnetic fields up to 15~T [Fig. \ref{fig:PM_B_Chis} (c),(d)] confirm the linear respectively non-linear behavior for the different magnetic field directions. For $B || a$ the slight non-linearity below $B^*$ is exhibited by a monotonous change for $T <$ \tn\ at applied magnetic fields $B \geq 3$~T as compared to $B = 1$~T. Overall, the data confirm spin-reorientation behaviour as for $B \geq 3$~T (i.e. above $B^*$) the susceptibility attains an almost constant value below \tn\ whereas it decreases sharply for $B || a = 1$~T. In addition, the data show the effect of external magnetic fields on the long-range spin ordered phase and particularly reveal stronger suppression of \tn\ for magnetic fields $B=15$~T applied in the $ab$ plane by $\Delta$\tn $= 18$~K as compared to the $B||c$ axis with yields $\Delta$\tn $= 2$~K. 

\begin{figure}[h]
\centering
\includegraphics[width=0.8\columnwidth]{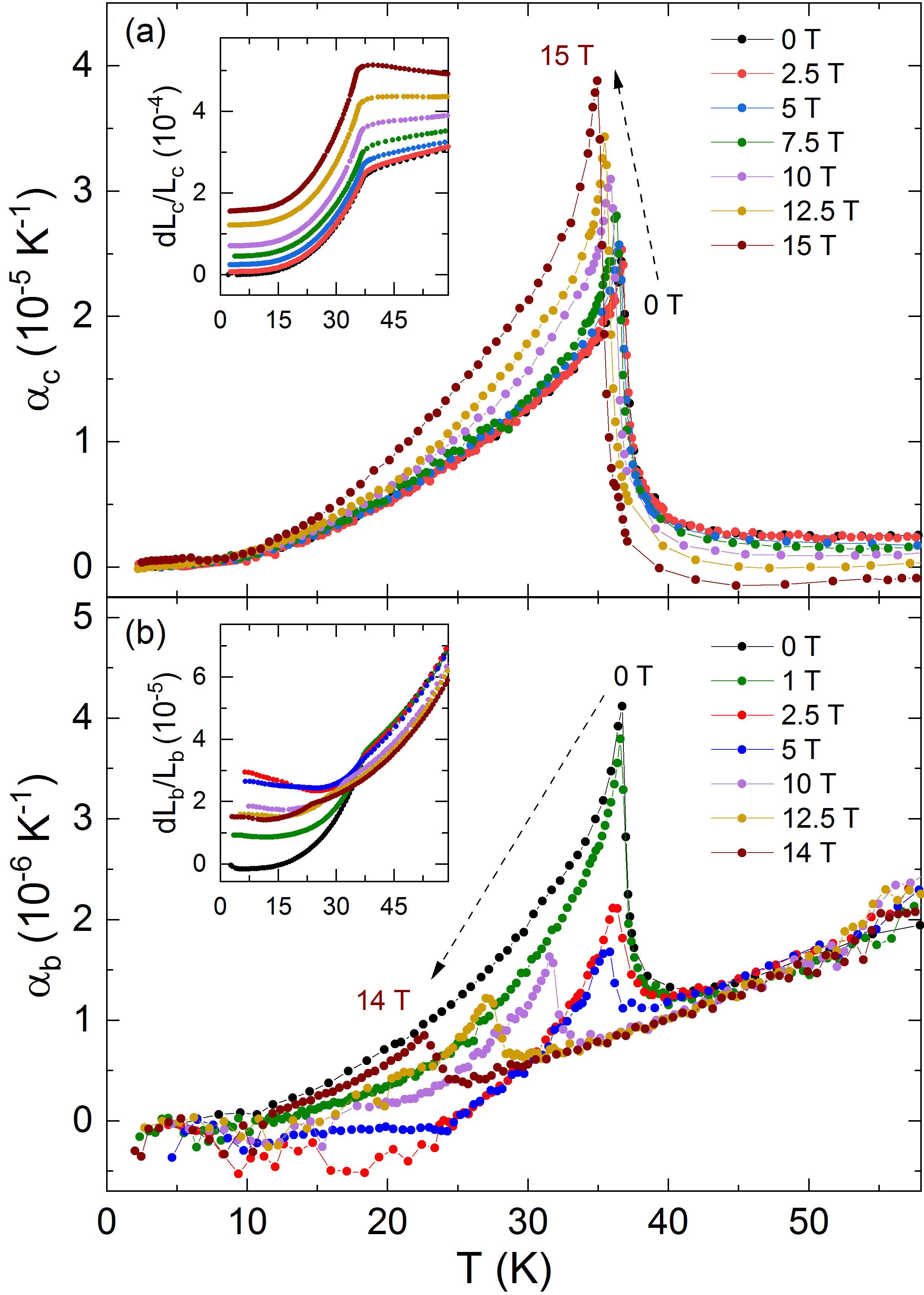}
\caption {Thermal expansion coefficients $\alpha_i$ at magnetic fields between 0 and 14~T magnetic fields applied along the (a) $c$- and (b) $b$ axes, respectively. The insets show the corresponding relative length changes $dL_i/L_i$ shifted by means of experimental magnetostriction curves, at $T$ = 50~K ($c$ axis) and 30~K ($b$ axis).}
\label{fig:TE_b_c_infields}
 \end{figure}

A strongly anisotropic field effect is also evident from Fig.~\ref{fig:TE_b_c_infields} where the thermal expansion coefficients $\alpha_i$ are shown for external magnetic fields up to 15~T. For $B || c$, \tn\ shifts slightly to lower temperatures along with a considerable increase in anomaly height on increasing magnetic fields. In contrast for $B || b$, suppression of the anomaly height is observed along-with a considerable shift of \tn\ to lower temperatures for fields above 5~T. The sharp anomalies in $\alpha_i$ facilitates construction of the magnetic phase diagram as shown in Fig.~\ref{fig:PD}. 

\begin{figure}[h]
\centering
\includegraphics[width=0.9\columnwidth]{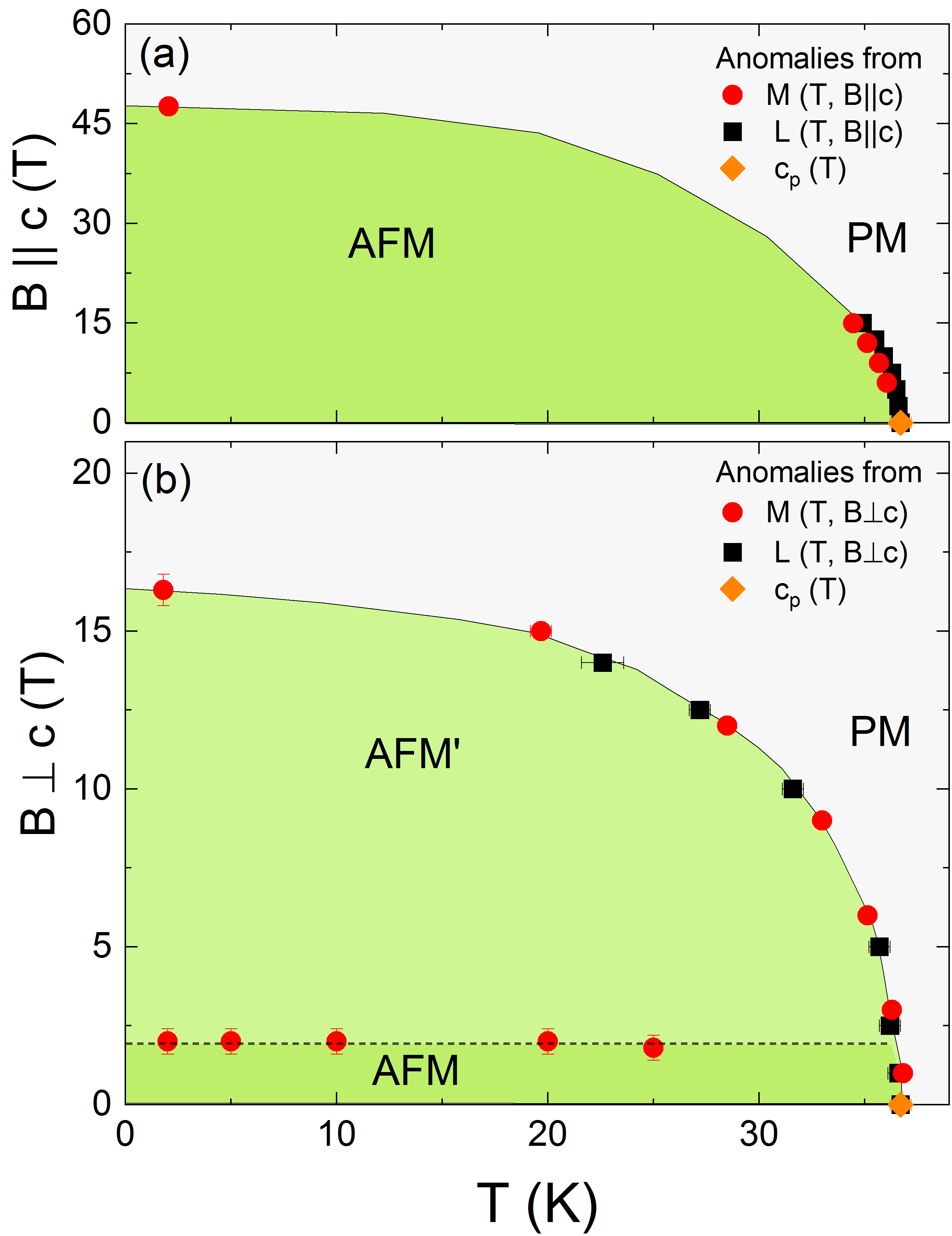}
\caption {Magnetic phase diagram of \cto\ for (a) $B||c$ axis and (b) $B||ab$ plane constructed from magnetization $M(T,B)$, dilatometry $L(T,B)$, and specific heat data. AFM, AFM' and PM label the antiferromagnetically ordered, spin-reoriented and paramagnetic region, respectively.}

\label{fig:PD}
 \end{figure}

\section{Discussion}

One of the intriguing properties of \cto\ is strong magnetoelectric coupling, the microscopic origin of which is still unclear. While our data show and elucidate pronounced magnetoelastic coupling, anomalies in the electrical permitivitty $\epsilon$ were observed at \tn\ for both polycrystalline~\cite{Harada2016} and single crystalline~\cite{Dubrovin_2020} \cto\ samples along with strong field-dependent magnetocapacitance at and above \tn\ ~\cite{Harada2016}. 

\begin{figure}[h]
\centering
\includegraphics[width=1 \columnwidth]{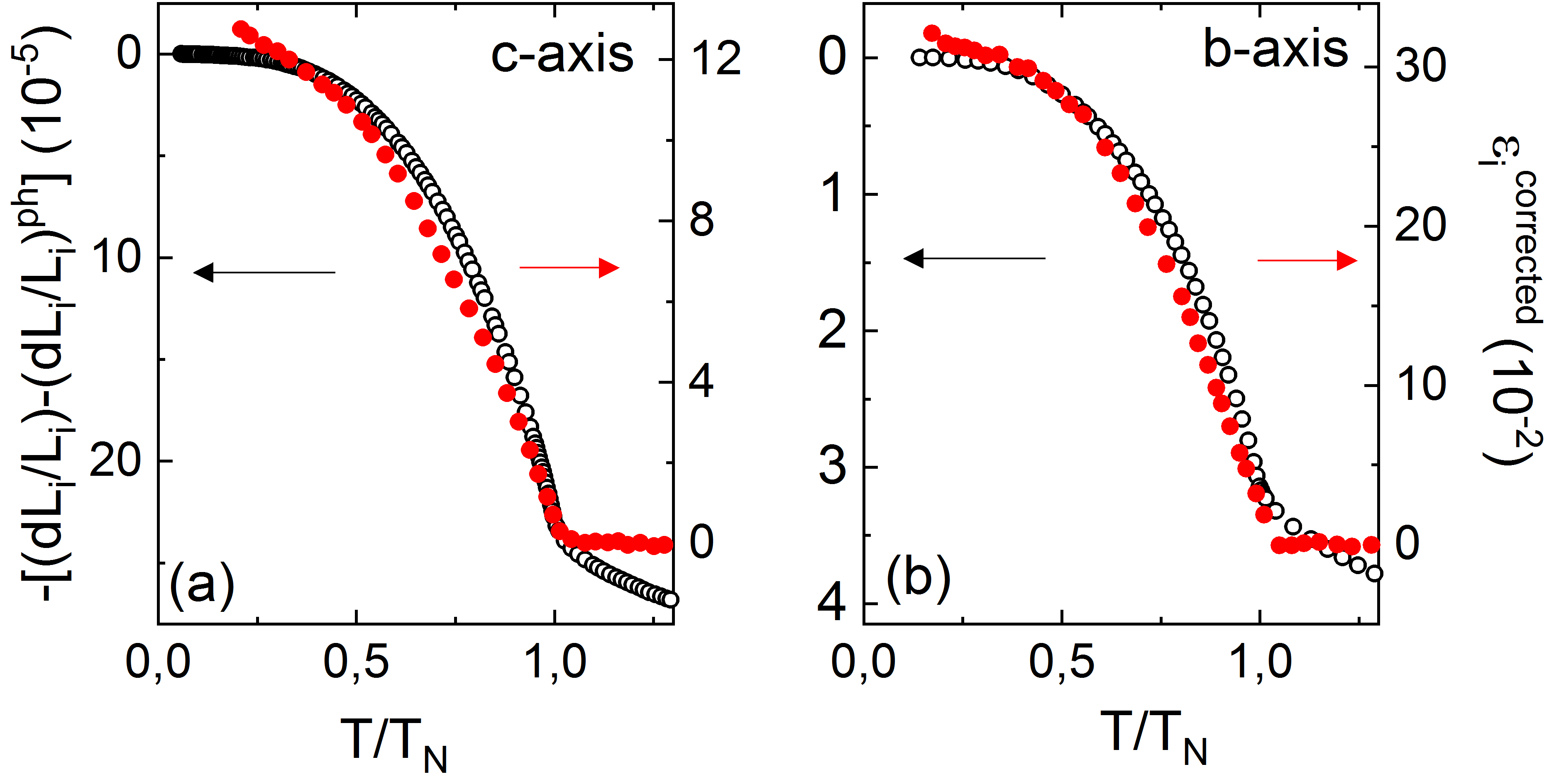}
\caption {Scalings of the non-phononic relative length changes $dL_i/L_i$ and corrected dielectric permitivitty $\epsilon$ digitized from Dubrovin \etal\ \cite{Dubrovin_2020} for (a) $c$ and (b) $b$ axis respectively.}
\label{fig:dllscaling_plot}
\end{figure} 

Comparison of the spontaneous strain, as measured by the relative length changes $dL_i/L_i$, with the dielectric permitivitty data published by Dubrovin \etal\ \cite{Dubrovin_2020}, sheds light on the coupling mechanism of lattice and dielectric degrees of freedom. As shown in Fig.~\ref{fig:dllscaling_plot}(b) and (c), the non-phononic relative length changes below \tn\ and the dielectric function $\epsilon$ show similar temperature dependence. Note that, the experimentally observed dielectric anomalies at \tn\ may also arise due to spontaneous deformation of lattice, as is exhibited by the relation $\epsilon = Cd/\epsilon_0A$, where $C, \epsilon_0, d$ and $A$ are sample capacitance, vacuum permitivitty, sample thickness and area, respectively. However, a direct quantitative comparison of $dL_i/L_i$ in Fig.~\ref{fig:dllscaling_plot}(a) and (b) with $\epsilon_i$ indicates that this cannot be the case. The relative length changes along $b$ and $c$ axes respectively are about four orders of magnitude smaller than the corresponding magnetic changes in $\epsilon_b$ and $\epsilon_c$, respectively. This observation indicates the presence of significant magnetodielectric coupling in \cto. The fact that the driving entropic changes at \tn\ are purely magnetic in nature as evidenced by Grüneisen analysis above furthermore implies that magnetic degrees of freedom form a single common origin for the observed structural, dielectric and magnetic changes at and below \tn.  

The effect of short-ranged magnetic correlations accompanied by a corresponding lattice response existing above \tn\ is evidenced from specific heat (Fig. \ref{fig:alpha_cp}(a)) and thermal expansion (Fig. \ref{fig:alpha_cp}(b)) measurements. It has been shown that $\bold{q}$-dependent spin-spin correlations couple to the dielectric response via the coupling of magnetic fluctuations to optical-phonons, thereby causing a significant magnetocapacitive effect \cite{Lawes_2003}. Accordingly, we conclude that the significant magnetocapacitive observed above \tn\ in polycrystalline \cto\ \cite{Harada2016} is due to persistent spin-spin correlations.   

Our results clearly show that both in-plane and out-of-plane uniaxial pressure enhance antiferromagnetism in \cto. Surprisingly, \nto\ which exhibits an easy-plane AFM structure\cite{dey_2021,Shirane_1959} similar to \cto , shows a different behaviour in associated length changes in the $ab$ plane when heating across \tn, i.e., an expansion along the $c$ axis implying d\tn/d$p_c > 0$ but shrinking of the $b$ axis implying d\tn/d$p_b < 0$ \cite{Dey2020}. In \nto, the opposing effects of in-plane and out-of-plane pressure are understood by enhanced and reduced strengths of the leading superexchange interactions for the respective uniaxial pressures \cite{Dey2020}.

In contrast for \cto, uniaxial pressures along both the in-plane and out of plane directions induce an increase of \tn. There are several potential explanations for this qualitatively different behaviour. Firstly, the electronic configuration of the Co$^{2+}$ ions in octahedral environment implies an effective orbital momentum of $l = 1$ as compared to the virtually quenched orbital momentum in Ni$^{2+}$ ions~\cite{Goodenough1967, Lines_1963}. Following the original work by Callen~\etal~\cite{Callen1968,Callen1965} and others~\cite{Frisch2012}, this difference which causes different anisotropy parameters can hence in principle result in opposite signs of magneto-elastic coupling coefficients. This is also demonstrated by magnetostriction studies on easy-plane antiferromagnets NiCl$_2$ and CoCl$_2$~\cite{Kalita_2005} which are isostructural the NiTiO$_3$ and CoTiO$_3$. Furthermore, our results are in-line with Ref.~\onlinecite{elliot2020}, where the presence of finite bond dependent magnetic frustration has been suggested for \cto. In this case, even small distortions are supposed to lift degeneracy resulting in large pressure effects~\cite{Kuchler_2017_frustration_lift}. Hence, frustration may be partially lifted by application of uniaxial pressure in the $ab$ plane generating additional effective in-plane anisotropy, thereby stabilizing magnetic order, i.e. d\tn/d$p_b > 0$.

We also note that the pressure dependencies of the magnetic anisotropy parameters influence the sign of the uniaxial thermal expansion coefficient (see equation (2) in \cite{Willemsen1977}). Therefore, one may speculate that a sufficiently large variation in the pressure dependencies of anisotropy could also lead to an opposite signs for $\alpha$ in the $ab$ plane in \nto\, and \cto.

\section{Conclusions}

To summarize, high-quality single crystals are used to refine the crystal structure and to study thermodynamic properties of \cto, which clearly demonstrate the presence of strong magnetoelastic coupling in \cto, and constructed its anisotropic magnetic phase diagram. By means of Grüneisen analysis we deduced the positive pressure dependencies of \tn\ for all crystallographic axes. We find that the magnetic degrees of freedom drive the observed structural, dielectric and magnetic changes both in the short-range ordered regime well-above \tn\ as well as in the long-range magnetically ordered phase.

\section{Acknowledgments}

This work has been performed in the frame of the International Max-Planck School IMPRS-QD. We acknowledge financial support by BMBF via the project SpinFun (13XP5088) and by Deutsche Forschungsgemeinschaft (DFG) under Germany’s Excellence Strategy EXC2181/1-390900948 (the Heidelberg STRUCTURES Excellence Cluster) and through project KL 1824/13-1. MAH. acknowledges the financial support from the Swedish Research Council (VR) under project No. 2018-05393. We acknowledge the support of the  HLD-HZDR, member of the European Magnetic Field Laboratory (EMFL).


\end{document}